%% file: main.tex
\documentclass[runningheads]{llncs}
\usepackage{tabularx}
\usepackage{booktabs}
\usepackage{graphicx}
\usepackage{subfigure}
\usepackage{cite}
\usepackage{multirow}
\usepackage{xcolor}
\usepackage{amsmath,amsfonts,amssymb}

\usepackage{marvosym}
\usepackage{pifont} 
\usepackage{mathtools}

\usepackage{color}

\usepackage{soul}

\usepackage{hyperref}
\let\svthefootnote\thefootnote
\newcommand\freefootnote[1]{%
  \let\thefootnote\relax%
  \footnotetext{#1}%
  \let\thefootnote\svthefootnote%
}

\usepackage{perpage} 
\MakePerPage{footnote} 

\begin{document}
%
\title{Knowledge-driven Subspace Fusion and Gradient Coordination for Multi-modal Learning}
\titlerunning{Subspace Fusion and Gradient Coordination for Multi-modal Learning}

\author{Yupei Zhang\inst{1}\textsuperscript{$*$}, Xiaofei Wang\inst{2}\textsuperscript{$*$}, Fangliangzi Meng\inst{3}, Jin Tang\inst{4}, Chao Li\inst{2,5,6}\textsuperscript{\Letter}}
\authorrunning{Y. Zhang et al.}

\institute{
Department of Pathology, The University of Hong Kong \and 
Department of Clinical Neurosciences, University of Cambridge, UK \and 
School of Life Sciences and Technology, Tongji University, China \and 
Zhejiang Lab, China \and
School of Science and Engineering, University of Dundee, UK \and
Department of Applied Mathematics and Theoretical Physics, University of
Cambridge, UK \\
\email{cl647@cam.ac.uk}
}
\maketitle              

\begin{abstract}
Multi-modal learning plays a crucial role in cancer diagnosis and prognosis. Current deep learning based multi-modal approaches are often limited by their abilities to model the complex correlations between genomics and histology data, addressing the intrinsic complexity of tumour ecosystem where both tumour and microenvironment contribute to malignancy. We propose a biologically interpretative and robust multi-modal learning framework to efficiently integrate histology images and genomics by decomposing the feature subspace of histology images and genomics, reflecting distinct tumour and microenvironment features. To enhance cross-modal interactions, we design a knowledge-driven subspace fusion scheme, consisting a cross-modal deformable attention module and a gene-guided consistency strategy. 
Additionally, in pursuit of dynamically optimizing the subspace knowledge, we further propose a novel gradient coordination learning strategy. Extensive experiments demonstrate the effectiveness of the proposed method, outperforming state-of-the-art techniques in three downstream tasks of glioma diagnosis, tumour grading, and survival analysis. Our code is available at \url{https://github.com/helenypzhang/Subspace-Multimodal-Learning}.
\freefootnote{$*$ Equal contribution.}
\end{abstract}
\section{Introduction}
\label{sec:introduction}

Multi-modal integration of genomics and histology becomes increasingly important for cancer diagnosis, evidenced by the recent shift of cancer taxonomy criteria to integrating molecular markers with histology features \cite{bombonati2011molecular,harris2010molecular,wang2023multi}.
However, joint analysis of multi-modal data at the clinic remains challenging \cite{wei2023multi}.
Automatic algorithms to effectively integrate multi-modal data of genomics and histology promise to offer rapid diagnosis and aid in precise cancer treatment.


Deep learning-based digital pathology  \cite{ chen2020pathomic, xing2022discrepancy} holds promise for rapid cancer diagnosis based on whole slide images (WSIs) derived from tissue sections\cite{zhang2022mutual}. Research \cite{lin2019ghrelin, wang2019mmp} has shown evidence of associations between genomics and WSIs, indicating that morphology features from WSIs may mirror genomic information. However, it remains a challenge to effectively integrate WSIs with genomic information for cancer diagnosis due to (1) the complexity of multi-modal data, i.e., WSIs in gigapixels and genomic profiles of tens of thousands genes and (2) the intrinsic tumour heterogeneity and complexity of cancers.



Previous studies have developed multi-modal approaches to integrate WSIs with genomics for cancer diagnosis. 
Among them, late-fusion methods integrate modality-specific features at the prediction layer. For instance, Chen \textit{et al.} \cite{chen2020pathomic} integrated genomics and WSIs using the Kronecker Product \cite{van2000ubiquitous} for cancer diagnosis and prognosis. However, late-fusion models ignore the interactions of the multi-modal data at their earlier learning stage, unable to  model the modality interaction for robust feature extraction and utilisation.

By contrast, intermediate-fusion techniques show promise to fuse modality-specific features at various levels before prediction. For instance, Chen \textit{et al.} \cite{chen2021multimodal} introduced a co-attention method that maps correlations between genomics and WSIs for survival analysis, enhancing survival predictions by learning dense co-attention mappings between genomics and bag representations of WSIs. Additionally, Zhou \textit{et al.} \cite{zhou2023cross} developed a multi-modal learning framework that delves into cross-modal correlations through inter-modality translation and alignment, offering complementary insights from different modalities for survival analysis.

Despite encouraging results, these models typically map WSIs to genomics in a singular embedding space, which may not fully reflect the complexity of cancer. In tumorigenesis, both tumour cells and tumor microenvironment\footnote{The tumor microenvironment is a complex ecosystem surrounding tumor cells, mainly composed of immune cells, also with other stromal cells and vessels.} (TME) contribute to malignancy \cite{li2019decoding}, providing essential insights with distinct morphological features for histology assessment. Further, evidence \cite{zeng2021exploration, zhou2021computational} shows that genetic profiles are associated with tumour and microenvironment characteristics. For example, isocitrate dehydrogenase (IDH) wildtype has marked necrosis and microvascular proliferation observed in WSIs. Therefore, decoding tumour- and TME-related morphology features from WSIs and genes from genomics promises to advance multi-modal integration. However, it remains challenging to effective model the regional interactions between WSIs and genomics. 

\begin{figure}[!t]
\includegraphics[width=\textwidth]{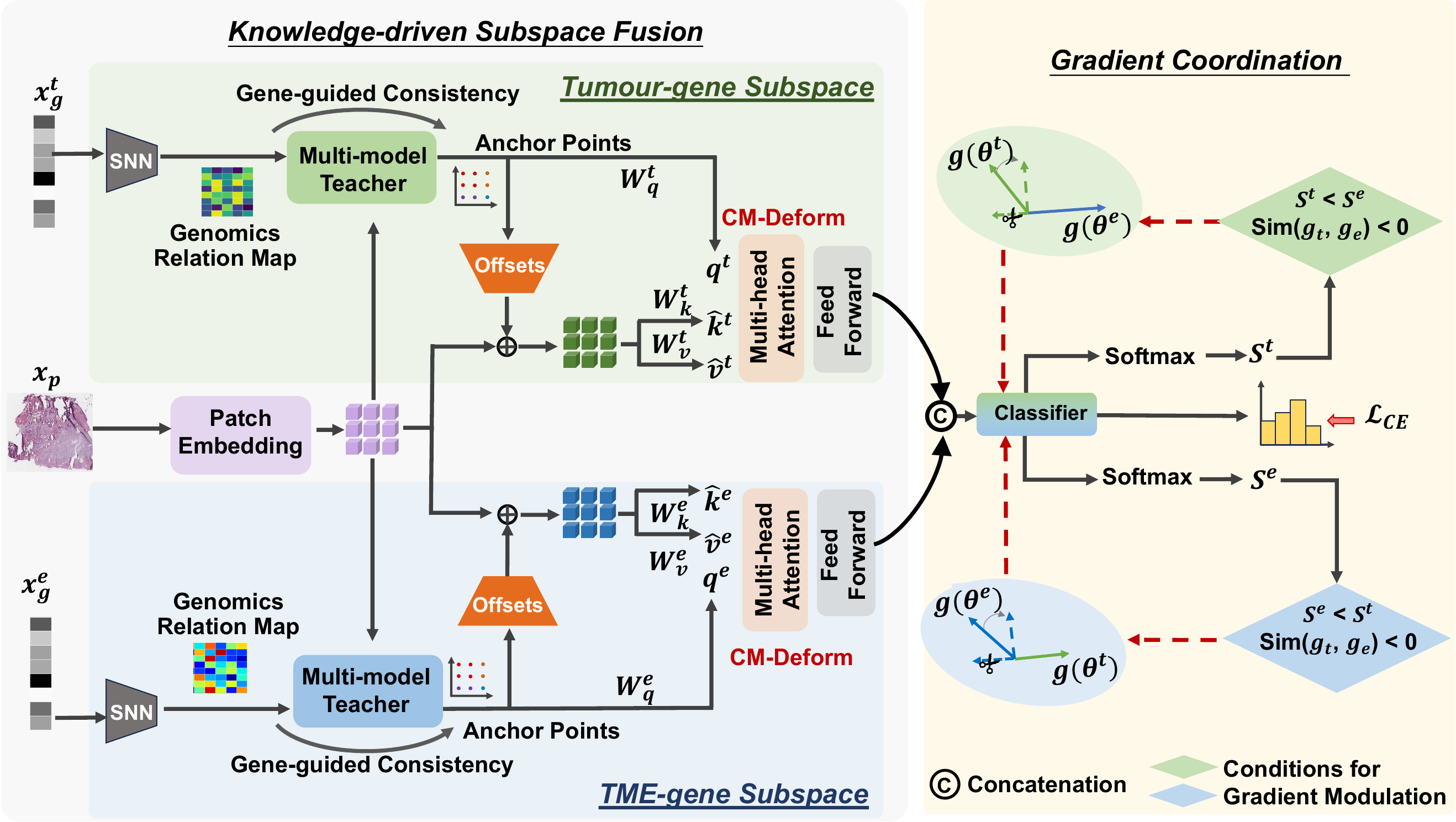}
\caption{Overview of the proposed framework. Left: Knowledge-driven subspace fusion scheme. Right: Confidence-guided Gradient Coordination. } \label{fig1}
\end{figure}

To tackle the challenge, We propose a novel multi-modal learning framework with a decoupled gene-to-histology integration strategy, for accurate and interpretable automatic cancer diagnosis. Specifically, our contribution is threefold:

\begin{itemize}
\item For the first time, we explicitly decompose the genomics data into tumour- and TME-related genes for effective integration with histological features.
\item  We propose a novel knowledge-driven subspace fusion (KS-Fusion) scheme to effectively enhance the multi-modal interactions. Specifically, in KS-Fusion, we present a cross-modal deformable attention (CM-Deform) module with a gene-guided consistency strategy to capture specific morphological features of the corresponding tumour- and TME-related genes. 
 \item We propose a confidence-guided gradient coordination (CG-Coord) scheme to regulate the multi-modal learning process, which promotes the optimal global performance through dynamic optimization.
\end{itemize}

Our extensive experiments on the three downstream tasks, i.e., cancer diagnosis, grading, and patient survival prediction, in two public datasets demonstrate that our
method outperforms other state-of-the-art (SOTA) methods.

\section{Methodology}
\subsection{Framework}
Fig.~\ref{fig1} illustrates the proposed multi-modal learning framework. Overall, to decompose the modal subspace features from genomics and WSIs and model their interactions, we propose a KS-Fusion scheme (left) and a CG-Coord scheme (right).
Specifically, as shown in the left part of Fig.~\ref{fig1}, categorized genomics features and histology features are first fused through a linear layer to generate a multi-modal teacher, which will be utilized as a query in the deformable attention module. To regulate the deformed offsets, we introduce a batch consistency to regulate the deformed offsets.
Then, as illustrated in the right part of Fig.~\ref{fig1}, the concatenated multi-modal feature is fed to the classifier for decision prediction, where the proposed CG-Coord scheme can alleviate the conflicts introduced by subspace gradients in training the classifier.

\subsection{Knowledge-driven Subspace Fusion}
Existing multi-modal fusion methods lack comprehensive modelling of feature subspaces for genomics profiles and WSIs morphological features. To enhance the biological guidance from niche-specific genomic profiles, we propose a KS-Fusion scheme to capture subspace informative features from both tumour- and TME-related gene profiles. Specifically, in the KS-Fusion scheme, the CM-Deform module along with the gene-guided consistency strategy is designated for efficiently extracting histological features of the corresponding tumour- and TME-related genes simultaneously.

\noindent\textbf{Cross-modal Deformable Attention Module.}
Overall, we build a two-stream neural network to model the cross-modal subspace interactions within 1) tumour-related genomics $x_g^t$ and the WSIs features $x_p$; 2) TME-related genomics $x_g^e$ and the WSIs features $x_p$, where $t$ signifies the tumour niche, $e$ denotes the TME niche, $g$ stands for genomics and $p$ represents WSIs.
Specifically, the genomic profile encoder is a spiking neural network (SNN)\cite{klambauer2017self} $\mathcal{G}$. 
The genomics data $x_g$ is divided into two groups, with $x_g^t$ representing the tumour-related genes and $x_g^e$ representing the TME-related genes. 

To realize the cross-modal deformable attention, we first apply a linear layer to obtain the multi-modal teacher features $x_{pg}^t$ and $x_{pg}^e$, which are then used to guide the offsets generation through an offsets generation network $\psi$, with two convolution layers and a scaler.
Original reference points are a uniform grid of points $p^u \in \mathbb{R}^{H_G \times W_G \times 2}$, given the input feature map $x_{pg}^u \in \mathbb{R}^{H \times W \times C}$, where $u$ represents the subspace and can be uniformly denoted as $u \in \{t, e\}$. For each stream, the multi-head cross-attention can be denoted as:
\begin{equation} \label{e:}
    q^u = x_{pg}^u W_{q}^u, \ \hat{k}^u = \hat{x}_p^u W_{k}^u, \ \hat{v}^u = \hat{x}_p^u W_{v}^u,
\end{equation}
where $q^u, \hat{k}^u, \hat{v}^u$ represent the query, deformed key, and deformed value, respectively the $W_q^u, W_k^u, W_v^u$ are projection networks. The $\hat{x}_p^u=F(x_p^u; \ {\rm norm}(p^u+\Delta p^u))$, $\Delta p^u = \psi(x_{pg}^u)$, and $F$ represents a bilinear interpolation function. For each stream, the output of an attention head is formulated as:
\begin{equation} \label{e:}
    z^{m;u} = {\rm softmax}(q^{(m);u}\hat{k}^{(m);u \top} / \sqrt{d})\hat{v}^{(m);u},
\end{equation}
where m represents the index of the attention head, with the range of 1 to M, and the final output is calculated as:
\begin{equation} \label{e:}
    z^u = {\rm concat}(z^{1;u}, ..., z^{M;u}) W_o^u,
\end{equation}
where $W_o^u$ is a projection network. In this way, the informative morphological features are dominated by the multi-modal features, enhancing the correlations and interactions between subspace genomic and histology embeddings.

\begin{figure}[!t]
\centering
\includegraphics[width=.8\textwidth]{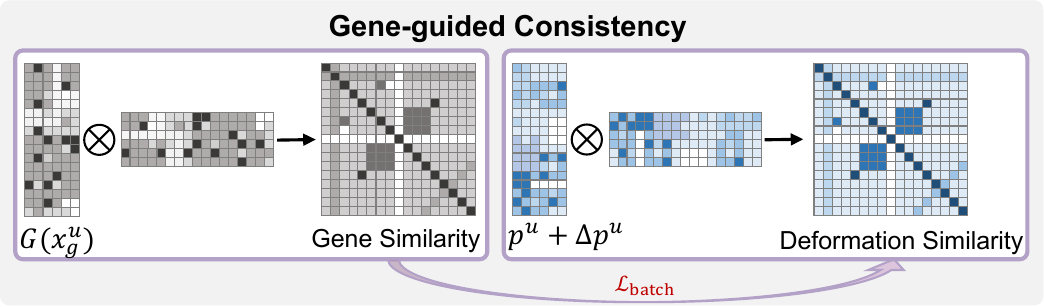}
    \caption{Illustration of gene-guided consistency strategy, which will regulate the deformation by the sample-wise similarity constraint.}
    \label{fig2}
\end{figure}

\subsubsection{Gene-guided consistency strategy.}
Furthermore, to ensure the intrinsic consistency between genomics and the deformed WSIs features, we introduce a batch consistency, denoted as gene-guided consistency strategy (Ge-Con), to further regulate offset adjustments in each subspace, which can be formulated as:
\begin{equation} \label{e:}
    \mathcal{L}^u_{\rm batch}\ = \ \frac{1}{B} ||\mathcal{S}_{b}(\mathcal{G}(x_{g}^u)) \ - \ \mathcal{S}_{b}(p^u+\Delta p^u)||_2,
\end{equation}
where $\mathcal{S}_b(x_1)$ denotes the Gram matrix of feature $x_1$, representing the correlations among individuals, and $B$ represents the batch size. As shown in Fig. \ref{fig2}, the sample-wise similarity further enhances the guidance of genomics. This gene-knowledge penetration towards WSIs features further enhances the subspace informative feature interaction and interaction, which ultimately facilitates multi-modal subspace fusion.

\subsection{Confidence-guided Gradient Coordination}

In training the classifier with multi-modal features, it is challenging to obtain a global optimum performance because the subspace gradients of tumour-gene and TME-gene may conflict when applying joint optimization. Therefore, to boost the downstream task by combining these two types of domain knowledge, we design the CG-Coord scheme to obtain the global training optimum via dynamic gradient regulation.

Specifically, assuming that $g(\theta^t)$ and $g(\theta^e)$ are in conflict when ${\rm cosine}(g(\theta^t), g(\theta^e))$ is smaller than zero, we adjust the gradient with less prediction confidence score when they conflict with each other. The adjustment process is formulated as:
\begin{equation} \label{e:coeffi}
    \begin{cases}
        \ \tilde{g}(\theta^t) \ = \ \Gamma (g(\theta^t), \ g(\theta^e)), & {\sum s^t < \sum s^e}, \\
        \ \tilde{g}(\theta^e) \ = \ \Gamma (g(\theta^e), \ g(\theta^t)), & {\sum s^e < \sum s^t},
    \end{cases}
\end{equation}
where $s^t = {\rm softmax}(\mathcal{D}(z^t))[k]$, $s^e = {\rm softmax}(\mathcal{D}(z^e))[k]$, and $\mathcal{D}$ is the decoder for the diagnosis and grading tasks. The confidence score for survival is represented by corresponding C-Index values. In equation \ref{e:coeffi}, $\Gamma(\vec{x}_1, \vec{x}_2)$ denotes prjection of the vector $\vec{x}_1$ to the perpendicular direction of vector $\vec{x}_2$. Besides, $\sum s^t$ and $\sum s^e$ represent the sum of the corresponding prediction scores on the k$-th$ label from tumour and TME branches in a mini-batch. In this way, gradients of different subspace features will be modulated to avoid conflicts dynamically, thereby achieving harmonious training.

\subsection{Training Objectives of Downstream Tasks}
In our framework, the training loss functions are tailored for varied downstream tasks, including cancer diagnosis, tumour grading, and prognosis prediction. Specifically, for the cancer diagnosis and grading tasks, we apply the cross-entropy loss as the task-specific objective, and the total training objectives can be formulated as:
\begin{align}
    \mathcal{L}_{\rm diag} &= L_{\rm CE}(\mathcal{D}(z^t, z^e; \ \theta_{diag}), \ Y_{\rm diag}) + \alpha \mathcal{L}_{batch}^t + (1-\alpha)\mathcal{L}_{batch}^e , \label{e:loss_diag}\\
    \mathcal{L}_{\rm grad} &= L_{\rm CE}(\mathcal{D}(z^t, z^e; \ \theta_{grad}), \ Y_{\rm grad}) + \alpha \mathcal{L}_{batch}^t + (1-\alpha)\mathcal{L}_{batch}^e , \label{e:loss_grad}
\end{align}
where $\mathcal{L}_{\rm CE}$ represents the cross-entropy loss, $\mathcal{D}$ represents a classifier, $\theta_{\rm diag}$ and $\theta_{\rm grad}$ represent the classifier parameters for diagnosis and grading tasks, respectively. $Y_{\rm diag}$ and $Y_{\rm grad}$ represent the diagnosis and grading labels, respectively. The $\alpha$ is a hyper-parameter for balancing subspace gene knowledge penetration. Notably, the hyper-parameter sensitivity analysis study can be found in Section 3.2.

Besides, for the prognosis prediction task, we adopt the NLL (negative log-likelihood) survival loss \cite{zhou2023cross}, denoted as $\mathcal{L}_{\rm NLL}$, as the task-specific objective for the survival outcome prediction, and the total training objective is:
\begin{equation} \label{e:loss_surv}
    \mathcal{L}_{\rm surv} = \mathcal{L}_{\rm NLL} + \alpha \mathcal{L}_{batch}^t + (1-\alpha)\mathcal{L}_{batch}^e ,
\end{equation}
In this way, our method can benefit downstream tasks by promoting subspace cross-modal fusion and harmonious training.

\section{Experiments \& Results}
\subsection{Datasets \& Implementation Details}
\noindent\textbf{Datasets}
We conduct experiments on two public datasets, i.e., TCGA GBM-LGG \cite{tomczak2015review} dataset and IvyGAP \cite{puchalski2018anatomic, shah2016data} dataset. The two datasets, focusing on gliomas, are merged as a meta dataset for better performance, including 2,387 tissue samples (668 cases) with paired WSIs and genomics profiles. We randomly split it into  training (534 cases), testing (68 cases), and validation (66 cases) sets.

WSIs are crop into patches sized at ${\rm224px} \times {\rm224px}$ of $0.5 \mu {\rm m \ px^{-1}}$. Following \cite{wang2023multi}, for each WSI, we extract 2500 patches with the biological repeat strategy.
For genomics process, according to previous studies \cite{bhattacharya2018immport}, we first sort the shared gene signatures in the TCGA GBM-LGG \cite{tomczak2015review} and IvyGAP \cite{puchalski2018anatomic, shah2016data} datasets according to the expression variance and then select the top 30\%, to capture important biological information in genes.
The genomic profile is enriched with 420 features, encompassing 59 tumour-related genes and 361 TME-related genes.

\noindent\textbf{Implementation Details}
All experiments were conducted using the PyTorch \cite{paszke2019pytorch} on two NVIDIA RTX A5000 GPUs, with a batch size of 8. Our method was trained for 20 epochs for diagnosis and grading, and 10 epochs for survival prediction. Network optimization was performed using the Adam optimizer \cite{kingma2014adam}.
The key hyper-parameters can be found in Supplementary Table I. Each hyperparameter was  tuned to achieve optimal performance on the validation set.

\subsection{Performance Evaluation}
\input{tables/table1}

\noindent\textbf{Baselines and SOTA Comparison Methods}
For each task, we compare our model with eight SOTA methods, with three uni-modal methods: AttMIL (Histology only)\cite{ilse2018attention}, TransMIL (Histology only) \cite{shao2021transmil}, SNN (Genomics only)\cite{klambauer2017self}) and five multi-modal fusion algorithms: Concat (AttMIL with SNN), Add (AttMILwith SNN), Bilinear (ResNet with SNN), MCAT \cite{chen2021multimodal}, and CMAT \cite{zhou2023cross}.

\begin{figure}[t!]
    \centering
\includegraphics[width=.6\textwidth]{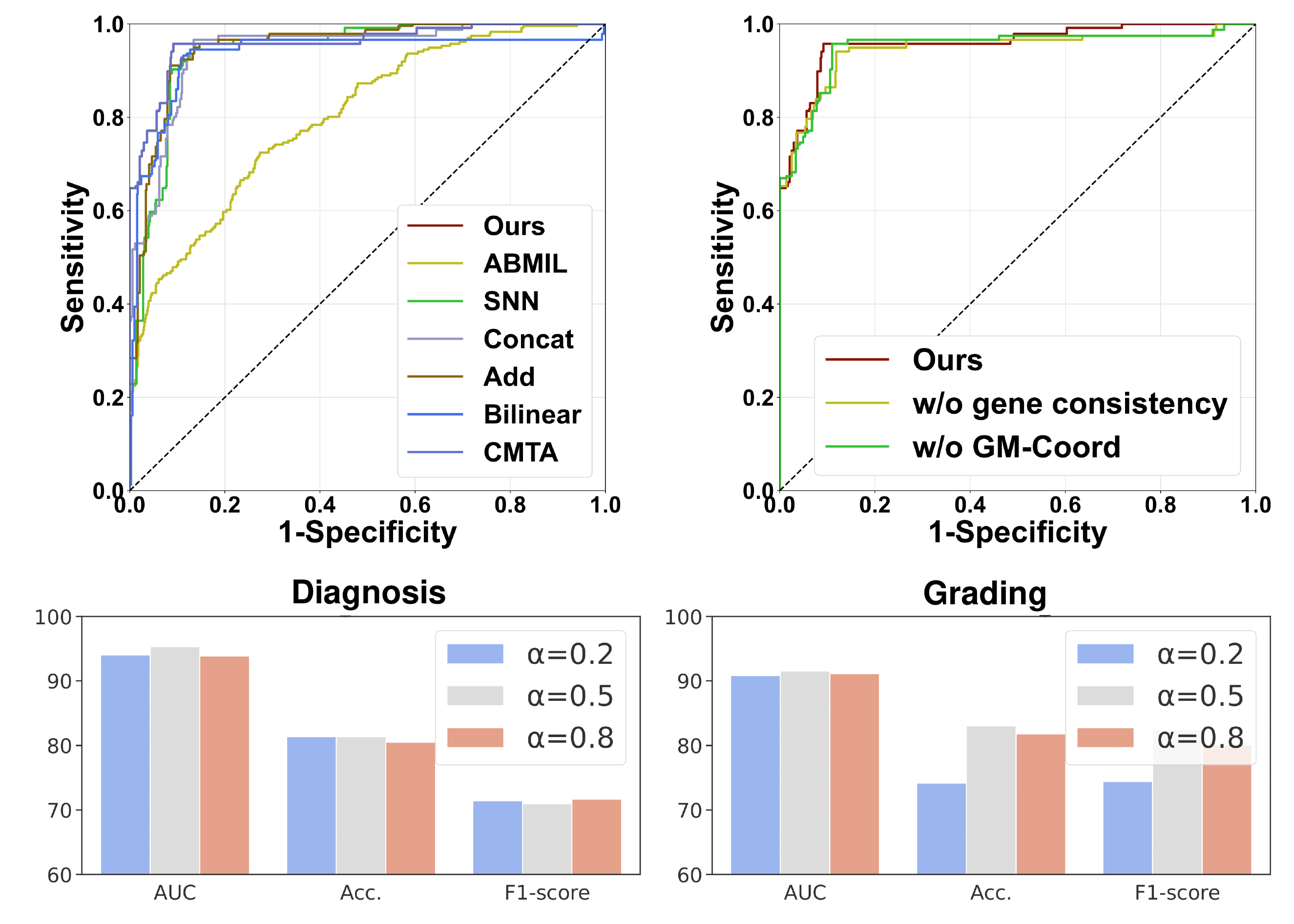}
    \caption{Top: ROCs of comparison and ablation study on glioma diagnosis task. Bottom: Hyper-parameter analysis of $\alpha$ in diagnosis and grading tasks.  }
    \label{fig:roc-alpha}
\end{figure}

\noindent\textbf{Downstream Task I: Glioma Diagnosis}
As shown in Table \ref{table1}, in gliomas diagnosis task, our framework is superior to all SOTA models, achieving  AUC of $95.28 \%$,  Acc. of $81.36 \%$,  Sen. of $72.92 \%$,  Spec. of $94.44 \%$, and  F1-score of $70.97 \%$. In terms of AUC and Acc., our framework outperforms others by at least $3.61 \%$ and $2.12 \%$, respectively, indicating the superior multi-modal learning ability of our method. Furthermore,  ROCs can be found in Fig. \ref{fig:roc-alpha} (top).

\noindent\textbf{Downstream Task II: Glioma Grading.} 
Gliomas can be categorized into four severity grades. As shown in the middle panel of Table \ref{table1}, our framework outperforms all SOTA models, achieving  AUC of $91.53 \%$, the Acc. of $83.05 \%$, the Sen. of $82.29 \%$, the Spec. of $91.29 \%$, and the F1-score of $82.43 \%$. In terms of AUC and F1-score, our framework outperforms others by at least $1.01 \%$ and $0.71 \%$, respectively, suggesting our excellent ability in the grading tasks.

\noindent\textbf{Downstream Task III: Survival Analysis} 
Following previous studies \cite{chen2020pathomic, chen2021multimodal}, we segment the overall survival time into four intervals based on the quartiles of event times of uncensored patients to compute the discretized-survival C-index. As shown in Table \ref{table1}, our framework outperforms state-of-the-art models, achieving the C-Index of $79.78 \%$, which outperforms the second-best method (marked in underline) by $2.74 \%$. The results show that our proposed multi-modal learning framework is also powerful in the prognosis prediction task.

\noindent\textbf{Hyper-parameters Sensitivity Analysis}
An additional hyper-parameter sensitivity analysis experiment is conducted to investigate the trade-off between tumour and TME subspaces in the proposed KS-Fusion scheme.
Specifically,  we conduct experiments on the hyperparameter $\alpha$ in Eq. \ref{e:loss_diag} and Eq. \ref{e:loss_grad}. 
As shown in Fig. \ref{fig:roc-alpha} (bottom), our method achieves the best performance when $\alpha$ equals 0.5, indicating that we can efficiently utilise both tumour- and TME-related gene features.

\noindent\textbf{Ablation Study}
\input{tables/table2}
To quantitatively evaluate the effectiveness of our proposed components, we conduct ablation studies for each component on two downstream tasks, i.e., cancer diagnosis and grading. In particular, we apply two ablative baselines of the proposed framework by disabling the gene-guided consistency strategy (denoted as $w/o \ {\rm Ge\text{-}Con}$) and Confidence-guided Gradient Coordination (denoted as $w/o \ {\rm CG\text{-}Coord}$). As illustrated in Table \ref{table2}, the AUC shows an improvement of $3.73 \%$ and $4.01 \%$ for the Ge-Con and CG-Coord correspondingly on the glioma diagnosis task. In terms of grading, the AUC increases $1.04 \%$ and $2.45 \%$ for Ge-Con and CG-Coord, respectively, indicating the effectiveness of our gene-guided consistency strategy and CG-Coord scheme.

\section{Conclusion}
Multi-modal data plays an increasingly important role in recent cancer diagnosis criteria. In order to effectively model the multi-modal data of histology and genomics, we propose a multi-modal learning framework with KS-fusion scheme, reflecting the intrinsic cancer mechanisms of tumour and TME.
In the KS-fusion scheme, we propose the CM-Deform module and gene-guided consistency strategy to enhance the multi-modal interaction among tumour-genes, TME-genes and histology images.  Besides, we also design a CG-Coord scheme to stabilize the multi-modal learning process, via dynamically adjusting the optimization of subspace features. Extensive experiments conducted on three  tasks demonstrate the effectiveness of our method, with a significant performance improvement compared to other SOTA methods, promising to promote precision oncology. 

%
%
%
\bibliographystyle{splncs04}
\bibliography{mybibliography}

\end{document}

%% file: tables/table1.tex
\begin{table*}[!t]
 \scriptsize
    \caption{Comparison with SOTA methods on three tasks. 
    p. and g. represent the modality of pathology and genomics, respectively.
    Best and second results are highlighted with \textbf{bold} and \underline{underline}.} \label{table1}
   \begin{center}
   \resizebox{.999\linewidth}{!}{ 
      \begin{tabular}{lcc | ccccc | ccccc | c}
         \toprule
         
         \multirow{2}{*}{Methods} 
         &\multirow{2}{*}{p.} 
         &\multirow{2}{*}{g.} 
         &\multicolumn{5}{c|}{Diagnosis, \%}   
         &\multicolumn{5}{c|}{Grading, \%} 
         &\multicolumn{1}{c}{Survival, \%}  \\
         
         \cmidrule(lr){4-14}  
         
          &       &       
          & AUC   & Acc.   & Sen.   & Spec.   &F1-score   
          & AUC   & Acc.   & Sen.   & Spec.   &F1-score
          & C-Index \\
          
         \midrule
         
         AttMIL~\cite{ilse2018attention}   
         &$\checkmark$     &   
         &80.25 &53.81 &38.77 &81.99 &33.30
         &79.70 &61.02 &61.36 &80.77 &60.30
         &55.59 \\
         
         TransMIL~\cite{shao2021transmil}   
         &$\checkmark$     &   
         &74.90 &51.69 &44.06 &84.11 &39.24 
         &83.14 &68.64 &67.69 &84.21 &66.83 
         &67.71 \\

         SNN~\cite{klambauer2017self}      
         &     & $\checkmark$  
         &90.87 &76.27 &64.79 &92.80 &64.94
         &\underline{90.52} &\textbf{83.05} &\underline{81.92} &\underline{91.18} &\underline{81.72}
         &\underline{77.04} \\

         Concat 
         &$\checkmark$     & $\checkmark$   
         &90.89 &74.58 &63.01 &92.35 &63.64 
         &89.32 &78.39 &76.76 &88.70 &75.00
         &75.06 \\
         
         Add       
         &$\checkmark$     & $\checkmark$  
         &\underline{91.67} &75.85 &63.90 &92.67 &63.56
         &90.36 &\underline{82.63} &81.60 &91.01 &81.39
         &73.42 \\

         Bilinear       
         &$\checkmark$     & $\checkmark$  
         &90.25 &\underline{79.24} &\underline{70.33} &\underline{93.50} &\underline{69.87}
         &82.49 &70.34 &68.27 &84.55 &62.82
         &73.71 \\
         
         MCAT~\cite{chen2021multimodal}                 
         &$\checkmark$     & $\checkmark$  
         &80.70 &55.08 &39.85 &84.68 &35.81
         &87.40 &59.75 &62.63 &80.54 &50.93
         &69.64 \\
         
         CMAT~\cite{zhou2023cross}
         &$\checkmark$     & $\checkmark$  
         &88.14 &68.64 &59.01 &89.87 &53.61
         &87.54 &54.66 &57.36 &78.07 &46.85
         &65.96 \\
         
         Ours      
         &$\checkmark$     & $\checkmark$  
         &\textbf{95.28} &\textbf{81.36} &\textbf{72.92} &\textbf{94.44} &\textbf{70.97}
         &\textbf{91.53} &\textbf{83.05} &\textbf{82.29} &\textbf{91.29} &\textbf{82.43}
         &\textbf{79.78} \\
         
         \bottomrule
      \end{tabular}
      }
   \end{center}
\end{table*}

%% file: tables/table2.tex
\begin{table*}[!t]
 \scriptsize
    \caption{Ablation studies on Diagnosis, Grading, and Survival tasks. The best results are highlighted with \textbf{bold}.} \label{table2}
   \begin{center}
   \resizebox{.999\linewidth}{!}{ 
      \begin{tabular}{l | ccccc | ccccc | c}
         \toprule
         
         \multirow{2}{*}{Methods} 
         &\multicolumn{5}{c|}{Diagnosis, \%}   
         &\multicolumn{5}{c}{Grading, \%}
         &Survival, \%
         \\
         
         \cmidrule(lr){2-12}  
            
          & AUC   & Acc.   & Sen.   & Spec.   &F1-score   
          & AUC   & Acc.   & Sen.   & Spec.   &F1-score& C-Index\\
          
         \midrule
         
         $w/o \ {\rm Ge\text{-}Con}$
         &91.55 &79.66 &69.43 &93.89 &67.07 
         &90.49 &71.61 &71.82 &85.77 &72.04 &72.16\\
         
         $w/o \ {\rm CG\text{-}Coord}$
         &91.27 &77.97 &67.47 &93.45 &67.70 
         &89.08 &73.73 &73.90 &86.82 &74.10 &76.54\\
         
         Ours
         &\textbf{95.28} &\textbf{81.36} &\textbf{72.92} &\textbf{94.44} &\textbf{70.97}
         &\textbf{91.53} &\textbf{83.05} &\textbf{82.29} &\textbf{91.29} &\textbf{82.43} &\textbf{79.78}\\
         
         \bottomrule
      \end{tabular}
    }
   \end{center}
\end{table*}